\documentclass[apjl]{emulateapj}
\usepackage{apjfonts}

\newcommand{\cmsq}{cm$^{-2}$}
\newcommand{\cmcc}{cm$^{-3}$}

\newcommand{\ergps}{ergs~s$^{-1}$}
\newcommand{\ergcms}{ergs~cm$^{-2}$~s$^{-1}$}

\newcommand{\cxc}{\textit{Chandra}}
\newcommand{\xmm}{\textit{XMM-Newton}}

\shorttitle{X-Ray Spectral Variability in V1118 Ori}
\shortauthors{Audard et al.}

\begin{document}


\title{X-Ray Spectral Variability During an Outburst in V1118 Ori}

\author{M.~Audard\altaffilmark{1}, M.~G\"udel\altaffilmark{2}, S.~L.~Skinner\altaffilmark{3}, K.~R.~Briggs\altaffilmark{2},
F.~M.~Walter\altaffilmark{4}, G.~Stringfellow\altaffilmark{3}, R.~T.~Hamilton\altaffilmark{5}, and
E.~F.~Guinan\altaffilmark{5}}

\altaffiltext{1}{Columbia Astrophysics Laboratory, Mail code 5247, 550 West
120$^\mathrm{th}$ Street, New York, NY 10027, {audard@astro.columbia.edu}}
\altaffiltext{2}{Paul Scherrer Institut, 5232 Villigen PSI, Switzerland, {briggs@astro.phys.ethz.ch, 
guedel@astro.phys.ethz.ch}}
\altaffiltext{3}{Center for Astrophysics and Space Astronomy, University of Colorado, 389 UCB, Boulder, CO 80309-0389, {skinners@casa.colorado.edu,
Guy.Stringfellow@colorado.edu}}
\altaffiltext{4}{Department of Physics and Astronomy, Stony Brook University, Stony Brook, NY 11794-3800, {fwalter@astro.sunysb.edu}}
\altaffiltext{5}{Department of Astronomy and Astrophysics, Villanova University, Villanova 19085, PA, {ryan.hamilton@villanova.edu, edward.guinan@villanova.edu}}

\begin{abstract}
We present results from a multi-wavelength campaign to monitor the 2005 outburst of the low-mass young star
\object{V1118 Ori}. Although our campaign covers the X-ray, optical, infrared, and radio regimes, we focus in
this Letter on the properties of the X-ray emission in V1118 Ori during the first few months after the optical outburst. 
\cxc\ and \xmm\  detected \object{V1118 Ori} at three epochs in early 2005. The X-ray flux and luminosity stayed 
similar within a factor of two, and at the same level as in a pre-outburst observation in 2002. The hydrogen column 
density showed no evidence for variation from its modest pre-outburst value of  $N_\mathrm{H} \sim 3 \times 10^{21}$~\cmsq.
However, a spectral change occurred from a dominant hot plasma ($\sim 25$~MK) in 2002 
and in January 2005 to a cooler plasma ($\sim 8$~MK) in February 2005 and in March 2005.  
We hypothesize that the hot magnetic loops high in the corona were disrupted by the closing in of the accretion disk 
due to the increased accretion rate during the outburst, whereas the lower cooler loops were probably less affected and 
became the dominant coronal component.

\end{abstract}


\section{Introduction}
\label{sect:intro}
The origin of X-rays remains an important topic for young accreting stars. Although
scaled-up, solar-like magnetic activity is the dominant mechanism in most
moderately accreting young stars (\citealt{gahm80,walter84,damiani95,damiani_micela95}, etc; see 
\citealt{bertout89,feigelson99,guedel04} for reviews), there is also
evidence that accretion may play an important role in others \citep{kastner02,stelzer04,schmitt05}.
Shocks in jets may also contribute to the soft X-ray component observed in some
young stars \citep{guedel05a,kastner05}.
Highly accreting stars can help us to understand how the accretion disk 
interacts with the stellar magnetosphere and photosphere and to understand the importance of accretion in the
X-ray emission in young stars.  A handful of such young stars display powerful eruptive events 
with flux increases in the optical regime of a few magnitudes. Such outbursts are thought to originate during a rapid 
increase of the disk accretion rate over a short period of time, from values of 
$10^{-7}$~$M_\odot$~yr$^{-1}$  to $10^{-4}$~$M_\odot$~yr$^{-1}$ \citep[see][]{hartmann96}.
The origin of the change in accretion rate is subject to debate, and could be triggered by
close companions or thermal disk instabilities. Two classes of erupting young stars have
emerged: FU Orionis stars (FUors), which display outbursts of 4 magnitudes and more that last
several decades, and EXors (named after the prototype EX Lup), which display smaller outbursts
($\Delta V=2-3$~mag) that last from a few months to a few years and that may occur repeatedly
\citep[e.g.,][]{herbig77}. 

On January 10, 2005, \citet{williams05} reported the outburst of \object{V1118 Ori}, 
a low-mass M1e young EXor star ($d = 470$~pc; $M_\star = 0.41 M_\sun$;
$R_\star=1.29 R_\sun$; $P_\mathrm{rot} = 2.23 \pm 0.04$~d; 
$L_\mathrm{bol}  \geq 0.25~L_\sun$; $\log T_\mathrm{eff} [K] = 3.562$; $\log t [yr] = 6.28$;
\citealt{hillenbrand97,stassun99}).
V1118 Ori has been known for its outbursts
in the past (e.g., 1983-84, 1988-90, 1992-94, 1997-98; see \citealt{garcia00} for details). 
For this 2005 outburst, we have started a  
multi-wavelength monitoring campaign. In this Letter, we report first results from our campaign, and 
focus on the observations in the X-ray regime obtained with \cxc\ and \xmm.

\section{Observations and Data Reduction}
\label{sect:data}

\subsection{X-Ray}

About two weeks after the IAU Circular by  \citet{williams05},  \facility{\textit{Chandra}} 
observed V1118 Ori, and \facility{\textit{XMM-Newton}} subsequently observed in February and March 2005.
We also retrieved archival data of serendipitous, pre-outburst observations of V1118 Ori by \xmm\  in
October 2001 and by \cxc\  in September 2002. The latter was published by \citet{ramirez04}
who reported the detection of V1118 Ori (source S029).
Table~\ref{tab:xray} provides a log of the observations.

The \cxc\  data were reduced with the CIAO 3.2.1 software with CALDB 3.01. 
For the 2005 observation, we extracted 12 events in the $0.1-10$~keV range from a circle 
with a radius of 2\arcsec\  centered at the position of V1118 Ori on the ACIS-S detector.
For the background, we used a nearby 10\arcsec\  radius circle. 
The scaled background contribution was 0.52~events. For the 2002 observation, in contrast, 
we used extraction radii of 15\arcsec\  and 45\arcsec\  for V1118 Ori and the nearby background, 
respectively, because of the large off-axis angle ($\theta = 8\farcm 85$) of V1118 Ori on the ACIS-I detector.

The \xmm\  data were reduced with the SAS 6.1.0 software. 
For the 2005 observations, events were extracted using circles with radii of 20\arcsec\  for V1118 Ori 
and 60\arcsec\  for the background located in a nearby source-free region. For the 2001 observation,
when only MOS2 data were available\footnote{The EPIC pn was off and the orientation of the MOS1 detector was such that 
V1118 Ori fell outside the field-of-view.}, we used circles of similar sizes as above at the expected position of 
V1118 Ori and for the background. Although ellipses would better represent the shape of the point-spread-function at 
large off-axis angles where V1118 Ori fell on the MOS2 detector, SAS does not accurately 
calculate effective areas for such shapes. 

\subsection{Optical and Near-Infrared}

We obtained optical ($V$, $R$, and $I$) and near-infrared ($J$, $H$, and $K$) photometry using 
the SMARTS\footnote{SMARTS, the Small and Medium Aperture Research Telescope Facility, is
a consortium of universities and research institutions that operate the
small telescopes at Cerro Tololo under contract with AURA.} 1.0~m and 1.3~m telescopes.
Magnitudes are measured differentially with respect to field stars.
We have used 2MASS magnitudes to convert the $JHK$ differential magnitudes to true observed magnitudes. 
For the optical bands, we have used arbitrary reference magnitudes of $V_\mathrm{ref} = 10.25$, 
$R_\mathrm{ref} = 9.5$, and $I_\mathrm{ref} = 8.75$  to match the magnitudes obtained by the Villanova group. 
Details about the SMARTS data will be provided elsewhere together with accompanying spectra
\citetext{Stringfellow et al.~2005, in preparation}.

Photometric coverage of V1118 Ori from the Villanova University Observatory 
was obtained from 2005 January 28 to March 16. Observations were 
carried out in standard Bessel $V$, $R$, and $I$ filters.  Dark and flat field frames
were collected at the end of each night's observations. We assigned a standard 
error of 0.05 mag for these observations, estimated from the signal-to-noise ratio and seeing 
conditions.

\subsection{Radio}
The \facility{{\em Very Large Array}} (VLA\footnote{The National Radio Astronomy Observatory (NRAO) is 
a facility of the National Science Foundation operated
under cooperative agreement by Associated Universities Inc.}) observed V1118 Ori 
on 2005 January 24 from 04:56 UT to 05:56 UT in hybrid BnA configuration,
using the continuum mode at 8.435~GHz (3.56~cm) with a total bandwidth of 100~MHz.
The primary and secondary calibrators were 3C48
and  J0541$-$056, respectively. The total on-source time was 
38 minutes for V1118 Ori.

Data were edited and calibrated using AIPS\footnote{Astronomical Image Processing System (AIPS)
is a software package developed by NRAO.}. 
We obtain a $3\sigma$ upper limit to the flux density of V1118 Ori
of $S_{3.6} \leq 0.075$~mJy, corresponding to $L_{3.6}$ $\leq$
$2 \times 10^{16}$~ergs~Hz$^{-1}$~s$^{-1}$.

\section{Results}
\label{sect:results}

Figure~\ref{fig:fig1} shows light curves of V1118 Ori in the optical and near-infrared bands, and in X-rays. 
During each of the X-ray observations, the X-ray flux stayed relatively constant and showed no strong variability 
due to, e.g., flares. We have thus used the full duration of the observations
for our spectral analysis. We fitted the background-subtracted spectra with 1-$T$ collisional 
ionization equilibrium models (additional components were not needed) based on the Astrophysical Plasma Emission Code 
\citep[APEC 1.3.1;][]{smith01} as implemented in XSPEC \citep{arnaud96} with a photoelectric 
absorption component, $N_\mathrm{H}$, generally free to vary and that uses cross sections from \citet{balucinska92}. Abundances
are relative to the solar photospheric values \citep{grevesse98}.
Table~\ref{tab:xray} summarizes the best-fit models, X-ray fluxes and luminosities with 68\% confidence
ranges.

The observations in 2001 and 2002 are useful to determine the pre-outburst X-ray properties
of V1118 Ori and compare them with those in the early phases of outburst in 2005.
Unfortunately, V1118 Ori was not detected in 2001 down to $F_\mathrm{X} < 2.3 \times
10^{-14}$~\ergcms\  (95\% Bayesian upper limit; \citealt{kraft91}). However, it was detected
in 2002 \citep[see][]{ramirez04} and its X-ray spectrum was dominated by a hot plasma 
($25$~MK) and low $N_\mathrm{H}$ ($3 \times 10^{21}$~\cmsq) after we fixed the metallicity at the outburst value, 
$Z=0.17$ (see Tab.~\ref{tab:xray}).

The \cxc\  January 2005 observation collected too few counts to fit a conventional binned spectrum.
We have instead used a maximum-likelihood fit method for unbinned 
data as described in \citet{guedel05b}. Briefly, the method compares the distribution of the 
event energies with the spectral energy distribution of template spectra characterized by 
the plasma $T$, $N_\mathrm{H}$, and $L_\mathrm{X}$, for a fixed metallicity 
(we adopted $Z=0.17$).  We determined that the plasma 
in V1118 Ori was hot ($>18$~MK), similar to the pre-outburst 2002 data. 

We fitted the \xmm\  EPIC pn and MOS spectra of February 2005 simultaneously. Whereas the X-ray flux and 
$N_\mathrm{H}$ did not vary much compared to January 2005 and September 2002, the X-ray spectrum of V1118 Ori was 
instead dominated by a cooler $8$~MK plasma with $Z = 0.17$. To establish whether a real change in plasma temperature 
and $N_\mathrm{H}$ occurred between 2002 and February 2005, we performed
simultaneous fits of these spectra, with each parameter (i.e., $T$ and
$N_\mathrm{H}$) in turn 
(a) tied to be the same value in both observations but free to vary, and (b) allowed to vary independently. We kept $Z$ fixed to 0.17
but left the normalizations free to vary in each observation. $F$-test results confirmed that the joint fit 
was significantly improved by allowing the temperature to be different ($F=27.59$, for a probability of $< 0.001$ 
that the improvement is by chance). On the other hand, allowing different $N_\mathrm{H}$ for a model with two temperature
components was not significantly better ($F=0.42$, for a probability of $0.52$). Figure~\ref{fig:fig2} shows the observed (absorbed) 
X-ray fluxed spectra in the hard 2002 pre-outburst state and in the February 2005 soft outburst state.

The March 2005 \xmm\ data were of lower quality than those of February 2005, due to a much higher background level 
resulting from solar activity. The pn data were more affected than the MOS data, so we fitted only the MOS1 and MOS2 
spectra simultaneously. We obtained cool temperatures of $6-16$~MK for plasma abundances between $Z=0$ (best-fit
abundance) and $Z=0.17$ (fixed value of February 2005). As above, we performed simultaneous fits of the March 
and February 2005 spectra. $F$-test results confirmed that the fit quality did not improve when the temperature ($F = 0.56$ for 
a probability of $0.46$) or $N_\mathrm{H}$ ($F=0.85$  for a probability of $0.36$) were allowed to be different in the two 
observations. However, the same comparison of the March 2005 data with that of September 2002 found a significant improvement 
in fit quality when we allowed the temperatures to be different ($F = 6.64$ with a probability of $0.01$) although not the 
$N_\mathrm{H}$  ($F = 1.47$ with a probability of $0.23$). Thus, the March 2005 plasma temperature is significantly cooler than 
in 2002, while in line with the temperature observed in February 2005. In conclusion, at some point after the  initial outburst, a marked 
change in plasma temperature occurred, which has remained unchanged on a time scale of at least one month.

\section{Discussion}
\label{sect:discussion}

\begin{figure}
\centering
\includegraphics[angle=0,width=0.75\textwidth]{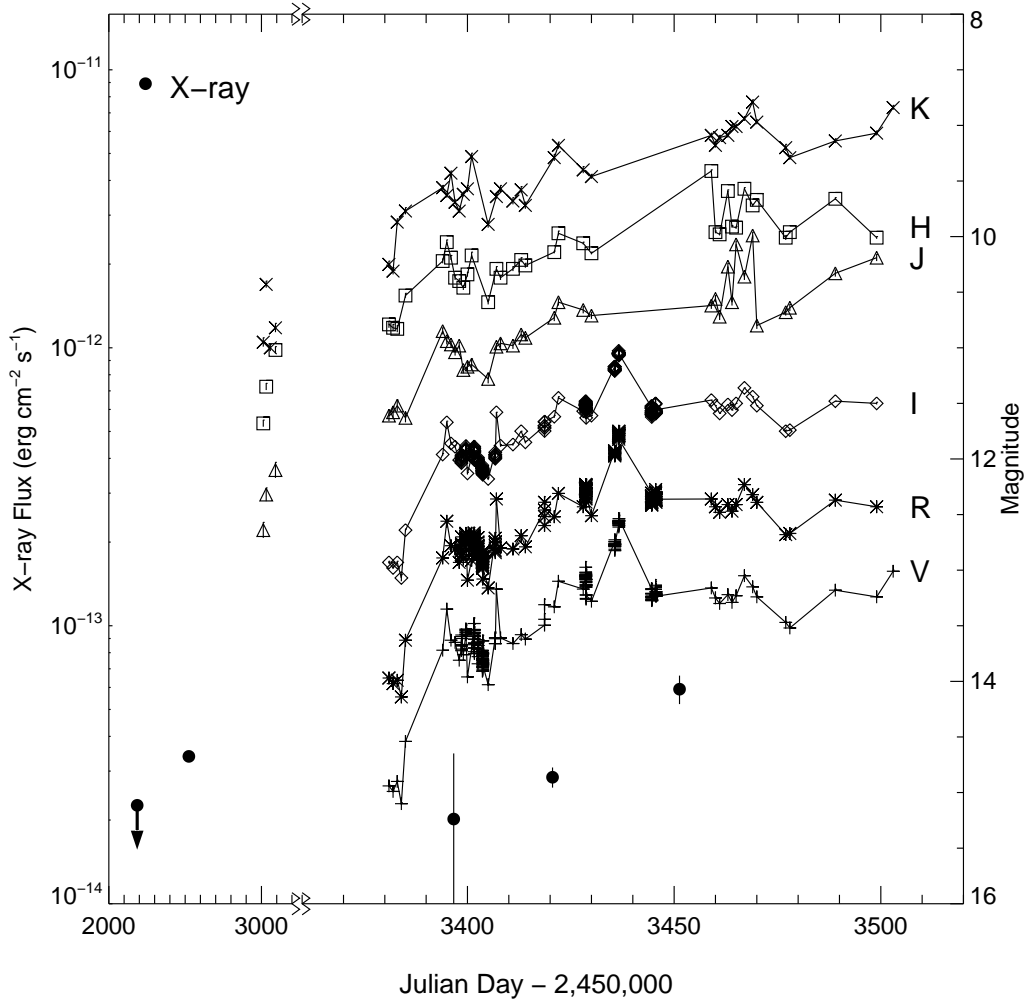}
\caption{Long-term light curves of V1118 Ori. Notice the broken time axis with different scales for
the pre-outburst epoch (1,200 days) and the outburst epoch (160 days). 
Constant zero-point magnitudes were added to the optical SMARTS data to eye-match the Villanova data (see text).
The uncertainties are plotted but are small compared to the day-to-day variability observed in the optical and near-infrared.
The left y-axis refers to X-ray data, whereas the right y-axis refers to the photometric optical and near-infrared 
data.\label{fig:fig1}}
\end{figure}

\begin{figure}
\centering
\includegraphics[angle=0,width=0.75\textwidth]{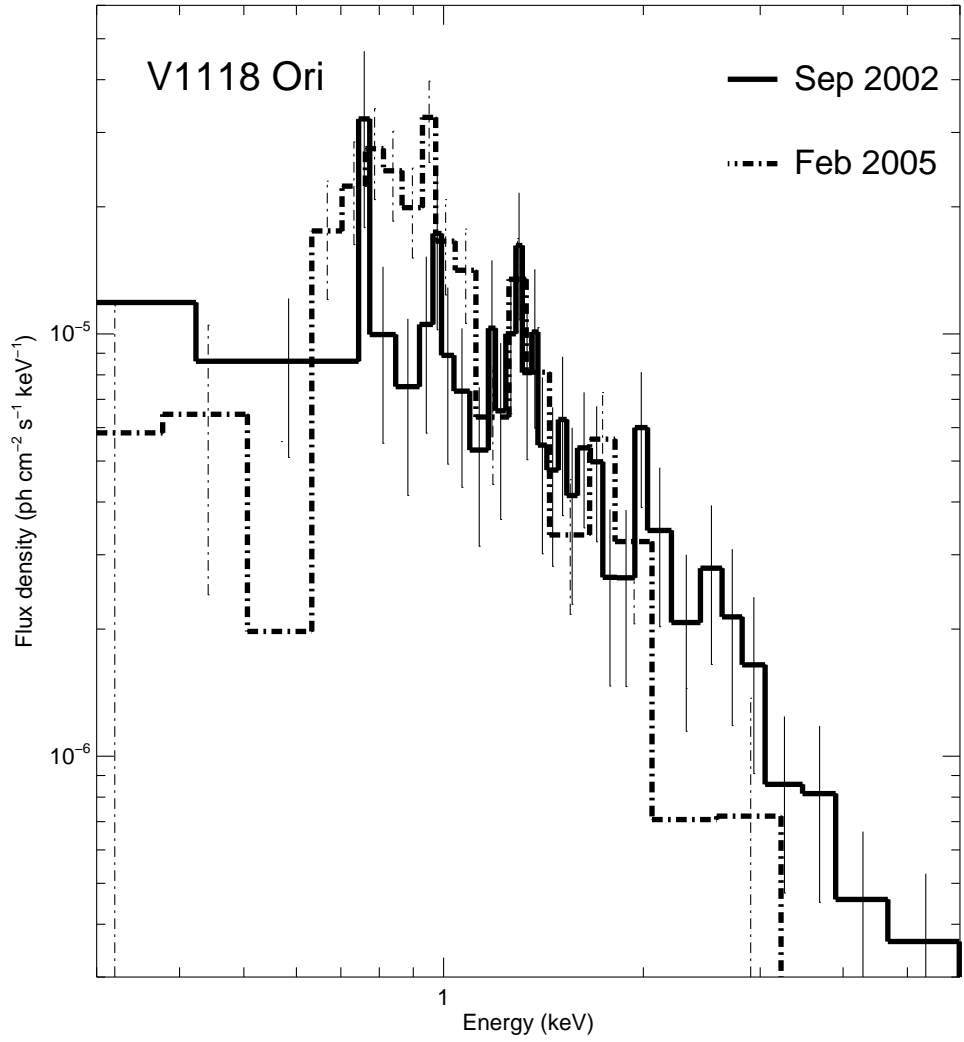}
\caption{Fluxed spectra (uncorrected for photoelectric absorption) of the September 2002 (solid) and February 2005 (dash-dotted) observations of V1118 Ori.
The soft excess ($0.6-1$~keV) is visible in the \xmm\  observation during the outburst. In contrast the hard ($> 2$~keV)
in the pre-outburst \cxc\  data is evident.
\label{fig:fig2}}
\end{figure}

The X-ray properties of the 2005 outburst of the EXor-type star V1118 Ori are remarkable in view of
its temperature change, which was not immediate but occurred a few months after the onset of the outburst 
in the optical and near-infrared regimes. Furthermore, the X-ray flux and luminosity of V1118 Ori varied by only
a factor of two during the outburst and compared to the serendipitous observation in 2002. The fluxes in the 
optical and near-infrared varied more significantly, within factors of $2-10$ (see Fig.~\ref{fig:fig1}).
The hydrogen column density stayed relatively low, $N_\mathrm{H} \sim (1-5) \times 10^{21}$~\cmsq,  
throughout the observations, giving no indication for additional absorption during the outburst. 

The disappearance of the hot plasma component, replaced by a dominant cool
plasma in February  and 
March 2005, is intriguing. There was no evidence that the high temperatures observed in 2002 and January 2005 were due
to flares. First, the X-ray light curves showed no evidence for flares. Second, flares in magnetically active
stars are accompanied by increases in both flux and luminosity. However, the X-ray fluxes, e.g., in 2002 ($F_\mathrm{X} \sim 
[3.4 \pm 0.3] \times 10^{-14}$~\ergcms)  and in February 2005 ($F_\mathrm{X} \sim 2.9^{+0.2}_{-0.3} \times 10^{-14}$~\ergcms) are similar. In addition,
the X-ray luminosity in February 2005 is about twice as large as the X-ray luminosity in 2002.
Thus, it is plausible that the dramatically increased accretion rate onto the star is responsible, 
directly or  indirectly, for the change in the plasma temperature.

When the accretion disk closed in near the star due to the increase in mass accretion rate, as
suggested by models for accreting binaries (\citealt{elsner77,ghosh79}), it plausibly disrupted the magnetosphere. 
Using the scaling laws of \citet{rosner78}, the height of a magnetic loop, $h$, is related to the temperature at its apex (most of the emission measure in a loop
has a similar temperature) and  the average electron density, $n_\mathrm{e}$ in \cmcc, as $h = 8.41 \times 10^5 T^2 / n_e$ in cm. 
If the loops have a density of $n_e$ is $10^{10}$~\cmcc, as is consistent with the densities observed in magnetically 
active stars \citep{ness04}, we get loop heights of $5 \times 10^{10}$ cm for $T=25$~MK, i.e., 
of the order of the stellar radius ($8.6 \times 10^{10}$~cm). In contrast, cooler loops (e.g., $8$~MK) have 
ten times smaller heights, i.e., they are smaller than the stellar radius and thus compact compared to the dimensions of the accretion disk; thus, 
when the accretion disk grows inward due to an increased mass accretion rate, the hotter loops interact first with the
accreting matter and are likely to be disrupted. A new magnetic configuration
then may have led preferentially to compact cool loops, thus enhancing the emission measure of the previously
undetected cool coronal component. 

The above properties contrast with those of another erupting young star, V1647 Ori, which displayed a
strong X-ray flux increase coincident with the optical/near-infrared outburst \citep{kastner04}.
In addition, V1647 Ori displayed a high $N_\mathrm{H}$ ($\sim 4 \times 10^{22}$~\cmsq) 
which was interpreted as an indication for a dense disk wind \citep{grosso05}. A cool
component was detected together with the hot component in V1647 Ori, but \citet{grosso05} interpreted the former
component as accretion shocks onto the stellar photosphere. It is unlikely that the cool component in V1118 Ori
originates from shocks because free-fall velocities of matter falling from the truncation radius are too slow. 
Even if X-rays  would be generated in shocks, they would likely be too absorbed to be detected \citep{drake05}.

The V1118 Ori observations suggest that the enhanced mass accretion rate during its outburst likely had an indirect
impact on the star's X-ray emission. Long-term multi-wavelength monitoring will provide important clues to understanding 
the importance of accretion in the X-ray emission in highly accreting young stars.

\acknowledgments

We thank the referee, J.~J. Drake, for useful comments and suggestions that improved this paper.
We acknowledge support by NASA through \cxc\ award DD5-6029X and through \xmm\ award NNG05GI96G
to Columbia University. The \cxc\ X-ray Observatory Center is operated by the Smithsonian Astrophysical 
Observatory for and on behalf of the NASA under contract NAS8-03060. Based on observations obtained with 
\xmm, an ESA science mission with instruments and contributions directly funded by ESA Member States and NASA. 
The PSI group acknowledges support from the Swiss National Science Foundation (grants 20-58827.99 
and 20-66875.01). Stony Brook's participation in SMARTS is made possible by support from the 
offices of the Provost and the Vice President for Research. We thank J. Allyn Smith, P. McGehee, 
J. Espinoza, and D. Gonzalez for doing the observations with the SMARTS telescopes.
We also thank H. Tannanbaum,  N. Schartel, and the VLA TOO panel for granting time to observe V1118 Ori, 
K. Arzner for providing us with the maximum-likelihood fit tools, and K.~Menou for stimulating discussions.


\clearpage

\begin{deluxetable}{lccccc}
\tabletypesize{\tiny}
\tablecolumns{6}
\tablewidth{0pc}
\tablecaption{Observation log and X-ray Properties of V1118 Ori.\label{tab:xray}}
\tablehead{
\colhead{\makebox[45mm][c]{Parameter}} 		&\colhead{Oct 2001}	& \colhead{Sep 2002} & \colhead{Jan 2005} & \colhead{Feb 2005} & \colhead{Mar 2005}}
\startdata
Satellite\dotfill				& \xmm		& \cxc				& \cxc		& \xmm			& \xmm 	\\
ObsID\dotfill					& \dataset[ADS/Sa.XMM#X/0112590301]{0112590301} 	& \dataset[ADS/Sa.CXO#obs/2548]{2548} & \dataset[ADS/Sa.CXO#obs/6204]{6204} & 
\dataset[ADS/Sa.XMM#X/0212480301]{0212480301} & \dataset[ADS/Sa.XMM#X/0212480401]{0212480401} \\
Duration (ks)\dotfill						& 40		& 48				& 5			& 20			  & 20				\\ 
Observation date (UT)\dotfill					& 2001 Oct 3 	& 2002 Sep 6--7			& 2005 Jan 26		& 2005 Feb 18-19	  & 2005 Mar 21			\\ 
\dotfill							& 01:02--12:03	& 12:57--02:54			& 03:07--05:04		& 22:36--04:50		  & 16:20--22:08		\\ 
Average Julian date - 2,450,000\dotfill				& 2,185.8	& 2,524.3			& 3,396.7		& 3,420.6		  & 3,451.3			\\ 
Net counts\dotfill						& $< 26.3$	& $169.7$			& $11.5$		& $494.1$\tablenotemark{a}& $252.5$\tablenotemark{a}	\\ 
$N_\mathrm{H}$ ($10^{21}$~\cmsq)\dotfill			& \nodata       & $2.7^{+1.2}_{-0.9}$		& $1.4^{+3.6}_{-1.4}$	& $4.3^{+1.3}_{-1.1}$     & $1.4^{+0.8}_{-0.5}$  	\\ 
$T$ (MK)\dotfill						& \nodata       & $25.1^{+6.3}_{-4.8}$		& $46^{+\infty}_{-28}$  & $7.7^{+1.3}_{-0.8}$     & $15.9^{+3.9}_{-2.0}$ 	\\ 
EM ($10^{53}$~\cmcc)\dotfill					& \nodata       & $1.5^{+0.4}_{-0.3}$		& $0.5^{+0.6}_{-0.2}$   & $4.0^{+1.4}_{-1.4}$     & $2.9^{+0.6}_{-0.5}$ 	\\ 
$Z/Z_\sun$\dotfill						& \nodata       & $:= 0.17$			& $:=0.17$       	& $0.17^{+0.20}_{-0.06}$  & $:=0.17$  			\\ 
$F_\mathrm{X}$ ($10^{-14}$~\ergcms)\tablenotemark{b}\dotfill	& $< 2.3$	& $3.1-3.7$			& $0.9-3.5$		& $2.6-3.1$		  & $5.2-6.6$			\\ 
$L_\mathrm{X}$ ($10^{30}$~\ergps)\tablenotemark{b}\dotfill	& $< 1.1$	& $1.5-1.8$			& $0.3-1.3$		& $3.2-3.8$		  & $2.5-3.2$			
\enddata
\tablecomments{The uncertainties are based on 68\% Bayesian confidence ranges, whereas the upper limits for 2001 are 95\% Bayesian upper limits.}
\tablenotetext{a}{Net counts for the sum of pn, MOS1, and MOS2 in February 2005 and for the sum of MOS1 and MOS2 in March 2005}
\tablenotetext{b}{X-ray luminosity and absorbed X-ray flux in the $0.1-10$~keV range, assuming $d=470$~pc.}
\end{deluxetable}


\clearpage


\clearpage

\end{document}